\documentclass[a4paper,11pt,amsmath,amssymb]{article} 
\usepackage{jheppub}
\usepackage{amsmath} 
\usepackage{array,relsize,float}
\usepackage{pstricks}
\usepackage{color} 
\usepackage{amssymb}
\usepackage{slashed}
\usepackage{tikz-cd}
\usepackage{graphicx} 
\usepackage{epsfig}
\usepackage{multicol} 
\usepackage{xspace}
\usepackage[T1]{fontenc}
\usepackage{slashed} 
\usepackage{hyperref}
\usepackage{pdfpages}
\usepackage{array}
\usepackage{tikz-cd}
\usepackage{amsfonts,layout,appendix,subfigure}
\allowdisplaybreaks
\input{epsf} 

\renewcommand{\thefigure}{\arabic{figure}}
 
\def\beq{\begin{equation}} \def\eeq{\end{equation}}
\def\beqn{\begin{eqnarray}} \def\eeqn{\end{eqnarray}}

\def\beq{\begin{equation}} 
\def\eeq{\end{equation}} 
\def\beqn{\begin{eqnarray}} 
\def\eeqn{\end{eqnarray}}

\def\nn{\nonumber}
\def\Eq#1{Eq.~(\ref{#1})}
\def\ii{\imath 0}

\def\qon#1{q_{#1,0}^{(+)}}

\def\qb{\mathbf{q}}

\pdfoutput=1

\newcommand{\valencia}{Instituto de F\'{\i}sica Corpuscular, Universitat de Val\`{e}ncia -- Consejo Superior de Investigaciones Cient\'{\i}ficas, Parc Cient\'{\i}fic, E-46980 Paterna, Valencia, Spain.}
\newcommand{\culiacanA}{Facultad  de  Ciencias de la Tierra y el Espacio,  Universidad  Aut\'onoma  de  Sinaloa,  Ciudad  Universitaria, CP 80000 Culiac\'an, M\'exico.}
\newcommand{\culiacanB}{Facultad  de  Ciencias  F\'isico-Matem\'aticas,  Universidad  Aut\'onoma  de  Sinaloa,  Ciudad  Universitaria, CP 80000 Culiac\'an, M\'exico.}
\newcommand{\salamanca}{Departamento de F\'isica Fundamental e IUFFyM, Universidad de Salamanca, E-37008 Salamanca, Spain.}
\newcommand{\europea}{Escuela de Ciencias, Ingenier\'ia y Diseño, Universidad Europea de Valencia, Paseo de la Alameda 7, 46010 Valencia, Spain.}

\begin{document}

\title{From five-loop scattering amplitudes to open trees with the Loop-Tree Duality}
\author[a,b,c]{Selomit Ram\'irez-Uribe,}
\author[c]{Roger J. Hern\'andez-Pinto,}
\author[a]{Germ\'an Rodrigo}
\author[d,e]{and German F. R. Sborlini}
\affiliation[a]{\valencia}
\affiliation[b]{\culiacanA}
\affiliation[c]{\culiacanB}
\affiliation[d]{\salamanca}
\affiliation[e]{\europea}

\emailAdd{selomit@ific.uv.es}
\emailAdd{roger@uas.edu.mx}
\emailAdd{german.rodrigo@csic.es}
\emailAdd{german.sborlini@usal.es}

\preprint{IFIC/22-30}

\abstract{Characterizing multiloop topologies is an important step towards developing novel methods at high perturbative orders in quantum field theory. 
In this article, we exploit the Loop-Tree Duality (LTD) formalism to analyse multiloop topologies that appear for the first time at five loops. Explicitly, we open the loops into connected trees and group them according to their topological properties. Then, we identify a kernel generator, the so-called N$^7$MLT \emph{universal topology}, that allow us to describe any scattering amplitude of up to five loops. Furthermore, we provide factorization and recursion relations that enable us to write these multiloop topologies in terms of simpler subtopologies, including several subsets of Feynman diagrams with an arbitrary number of loops. Our approach takes advantage of many symmetries present in the graphical description of the original fundamental five-loop topologies. The results obtained in this article might shed light into a more efficient determination of higher-order corrections to the running couplings, which are crucial in the current and future precision physics program.}

\setcounter{page}{1}
\maketitle

\section{Introduction and motivation}
\label{sec:introduction}
Nowadays, the state-of-the-art in particle physics moves around the challenge of breaking the precision frontier. As a new generation of experiments~\cite{Abada:2019lih,Abada:2019zxq,Benedikt:2018csr,Abada:2019ono,Bambade:2019fyw,Roloff:2018dqu,CEPCStudyGroup:2018ghi} is approaching and current colliders are collecting enormous amounts of data, there is an increasing pressure in the theory community to provide more precise predictions. Thus, it is mandatory to explore novel techniques to deal with the complex mathematical structures behind the Standard Model (SM), and Quantum Field Theories (QFT) in general.

Most of the existing phenomenologically-relevant gauge theories, with SM and QCD as flagship examples, are not exactly solvable with our current technologies. As a consequence, we rely on approximate methods, which are valid under certain assumptions. In the high-energy limit, it turns out that perturbation theory leads to the most reliable description of particle collisions, and provides a systematic approach to calculate precise predictions. 
This approach is based on series expansions involving Feynman diagrams, and reaching higher-orders requires to computing more and more loops. Beyond two loops, Feynman diagrams are hard to compute analytically and numerical methods quickly lose efficiency.

Even if several strategies were devised in the last decades~\cite{Heinrich:2020ybq}, the loop-calculation bottleneck is still a challenge. In this direction, the Loop-Tree Duality (LTD)~\cite{Catani:2008xa,Bierenbaum:2010cy,Bierenbaum:2012th,Buchta:2014dfa,Tomboulis:2017rvd,Runkel:2019yrs,Capatti:2019ypt,Verdugo:2020kzh,deJesusAguilera-Verdugo:2021mvg,Kromin:2022txz} emerges as a powerful approach to transform complex loop integrals into phase-space ones, which are more suitable for numerical calculations \cite{Buchta:2015wna,Buchta:2015xda,Jurado:2017xut,Driencourt-Mangin:2019yhu,Capatti:2019edf,Kermanschah:2021wbk}, asymptotic expansions \cite{Driencourt-Mangin:2017gop,Plenter:2019jyj,Plenter:2020lop,Plenter:2022zxk} and the local renormalization of ultraviolet singularities \cite{Driencourt-Mangin:2019aix,Driencourt-Mangin:2019sfl}. In particular, LTD leads to a more transparent interpretation of the singular IR-structure of multiloop integrals \cite{Buchta:2014dfa,Aguilera-Verdugo:2019kbz}, paving the road for the development of novel computational strategies. For instance, LTD is a key ingredient of the so-called Four-Dimensional Unsubtraction (FDU) ~\cite{Hernandez-Pinto:2015ysa,Sborlini:2016fcj,Sborlini:2016gbr,Sborlini:2016hat,Driencourt-Mangin:2019sfl}, a frameworks that aims to combine the real, virtual and renormalization counter-terms into a unique and integrable function in the four physical dimensions for the space-time, by-passing any additional regularization method. 

Still, LTD has another interesting feature: it manifestly exhibits the causal nature of Feynman diagrams and scattering amplitudes. By using the nested residue strategy \cite{Verdugo:2020kzh,Ramirez-Uribe:2020hes,Aguilera-Verdugo:2020nrp}, one degree of freedom per loop is removed. In particular, integrating out the energy component of each loop by applying Cauchy's residue theorem, we end up with a multi-dimensional integration in an Euclidean space. This leads to an integrand representation which is very compact and free of spurious unphysical singularities. Explicitly, adding all the terms generated by the nested residues, only those compatible with causality remain. 
The result is a representation that allows a direct interpretation in terms of causal entangled thresholds \cite{Verdugo:2020kzh,Aguilera-Verdugo:2020kzc,TorresBobadilla:2021dkq,Sborlini:2021owe,TorresBobadilla:2021ivx,snowmass2020,Capatti:2020ytd}, and enables a more stable numerical calculation. 

Besides, this novel LTD formulation allows a classification of families of Feynman diagrams according to the number of sets of propagators~\cite{Verdugo:2020kzh}. We call Maximal Loop Topology (MLT) to those $L$-loop diagrams with $L+1$ sets of propagators; Next-to-Maximal Loop Topology (NMLT), if they have $L+2$ sets of propagators; and so on. Interestingly, it was proven that factorization formulae exist~\cite{Ramirez-Uribe:2020hes,Aguilera-Verdugo:2020nrp}, and N$^k$MLT topologies can be reduced to convolutions of N$^{j}$MLT and MLT subtopologies (with $j<k$). Furthermore, the singular structure of the more complex topologies is constrained by these factorization properties, thus simplifying their treatment and cancellation.

The concrete purpose of this work is to study the application of the LTD approach to describe families of multiloop topologies that appear for the first time at five loops. As we did in Ref. \cite{Ramirez-Uribe:2020hes} with the analysis of four-loop topologies, we define here a N$^7$MLT \emph{universal topology} that allows to generate the LTD representation of any possible five-loop Feynman diagram or scattering amplitude. We exploit the recursive formulation in terms of simpler subtopologies and MLT insertions, in order to extract the main features of the more complex diagrams.

The outline of this paper is the following. In Sec. \ref{sec:LTD} we present a brief description of the LTD framework, with special emphasis in the nested residue strategy. Then, in Sec. \ref{sec:topology} we explain the construction of the \emph{universal} N$^7$MLT kernel generator. Subtopologies counting and relevant factorization formulae are provided, and we present a concrete application example in Sec. \ref{sec:Examples}. Finally, conclusions and possible future research directions are discussed in Sec. \ref{sec:conclusions}. 

\section{LTD framework}
\label{sec:LTD}
Any loop scattering amplitude is written in the Feynman representation as an integral in the Minkowski space of the $L$ loop momenta as,
\begin{align}\label{eq:integral}
\mathcal{A}^{(L)} = \int_{\ell_1, \ldots, \ell_L} \mathcal{A}_F^{(L)}(1,\ldots, n) \, ,
\end{align}
where the integration measure in dimensional regularization~\cite{Bollini:1972ui, tHooft:1972tcz} is given by 
\beq
\int_{\ell_s} = -\imath \mu^{4-d} \int \frac{d^d\ell_s}{(2\pi)^d}\, ,
\eeq
with $d$ the number of space-time dimensions and $\mu$ an arbitrary energy scale.
The integrand in \Eq{eq:integral} considers $N$ external legs, and is composed by the product of Feynman propagators and the numerator given by the specific theory,
\begin{align}\label{eq:amplitude}
&\mathcal{A}_F^{(L)}(1,\ldots, n)  = \mathcal{N}( \{ \ell_s\}_L,  \{ p_j\}_N) \, G_F(1,\ldots, n) \, ,
\end{align}
with
\begin{align}
G_F(1,\ldots, n) = \prod_{i\in 1\cup\ldots \cup n}  G_F(q_i) \, ,
\end{align}
standing for the product of Feynman propagators of one set that depends on a specific loop momentum or the union of several sets that depend on different linear combinations of the loop momenta. 
We write Feynman propagators in the following alternative form,
\begin{equation}
G_F(q_i) = \frac{1}{(q_{i,0}-\qon{i})(q_{i,0}+\qon{i})}~,
\label{eq:propagator}
\end{equation}
exhibiting the location of the poles, $\qon{i} = \sqrt{\qb_i^2+m_i^2-\ii}$, with $q_{i,0}$ and $\qb_i$ the temporal and spacial components of $q_i$, respectively, and $m_i$ the mass of the particle. 
It is important to remark that the internal structure of $\mathcal{A}_F^{(L)}$ is implicitly specified via the overall tagging of the different
sets of internal momenta.

To compute the LTD representation, we integrate out one of the components of the loop momenta by applying the Cauchy's residue theorem. 
Regarding multiloop scattering amplitudes, the LTD representation is expressed in terms of nested residues~\cite{Verdugo:2020kzh,Aguilera-Verdugo:2020nrp},
\begin{align}
\label{eq:nested}
&\mathcal{A}_D^{(L)}(1,\ldots, r; r+1,\dots, n)  
=-2\pi \imath \sum_{i_r \in r} {\rm Res} (\mathcal{A}_D^{(L)}(1, \ldots, r-1;r, \ldots, n), {\rm Im}(\eta\cdot q_{i_r})<0)\, , 
\end{align}
starting from 
\begin{align}
\mathcal{A}_D^{(L)}(1; 2, \ldots, n)
=-2\pi \imath \sum_{i_1 \in 1} {\rm Res} (\mathcal{A}_F^{(L)}(1, \dots, n), {\rm Im}(\eta\cdot q_{i_1})<0)\, , 
\end{align}
where $\mathcal{A}_F^{(L)}(1, \dots, n)$ is the integrand in the Feynman representation in Eq.~\eqref{eq:integral}. 
The sets located before the semicolon have one on-shell propagator, whereas the sets appearing after the semicolon contain off-shell propagators only. 
The residues are evaluated through the selection of the poles with negative imaginary components by implementing the future-like vector $\eta$ selecting the component of the loop momenta to be integrated. 
We conveniently take $\eta^{\mu}=(1,{\bf 0})$, allowing to work in the integration domain of the loop three-momenta of an Euclidean space instead of a Minkowsky space.  

\section{Universal topology}
\label{sec:topology}
In this section we analyse the multiloop topologies that appear for the first time at five loops, i.e. the family composed by those loop topologies described by $L+5$, $L+6$ and $L+7$ propagators.
We start identifying the representative diagrams, followed by a unified description and the application of the LTD framework to obtain the dual opening to connected trees.

Based on the classification scheme presented in Refs.~\cite{Verdugo:2020kzh,Aguilera-Verdugo:2020nrp}, the topological complexity of this set of diagrams corresponds to N$^7$MLT, which has $L+5$ common sets of propagators:
\beq
\{1, \ldots, L+1, 12, 123, 1234, 2345\}~, 
\eeq
and two additional specific sets distinguishing them from each other. 
To achieve a global representation of this family, we first need to unravel the representative topologies through the identification of the distinctive pair of sets. 
\begin{figure}[t]
\includegraphics[scale=0.38]{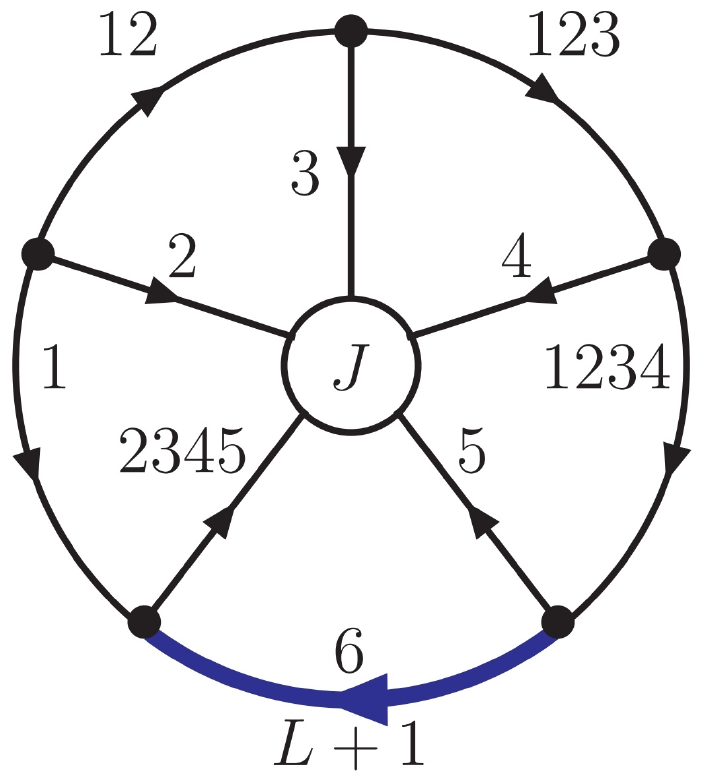}
\centering
\caption{Diagrammatic representation of the general N$^7$MLT topology. The current $J$ considers all the possible combinations among the internal propagators. The blue line encodes a MLT with $L-5$ loops and $L-4$ propagators. 
\label{fig:topologies5_J}}
\end{figure}

\begin{figure}[t]
\includegraphics[scale=0.45]{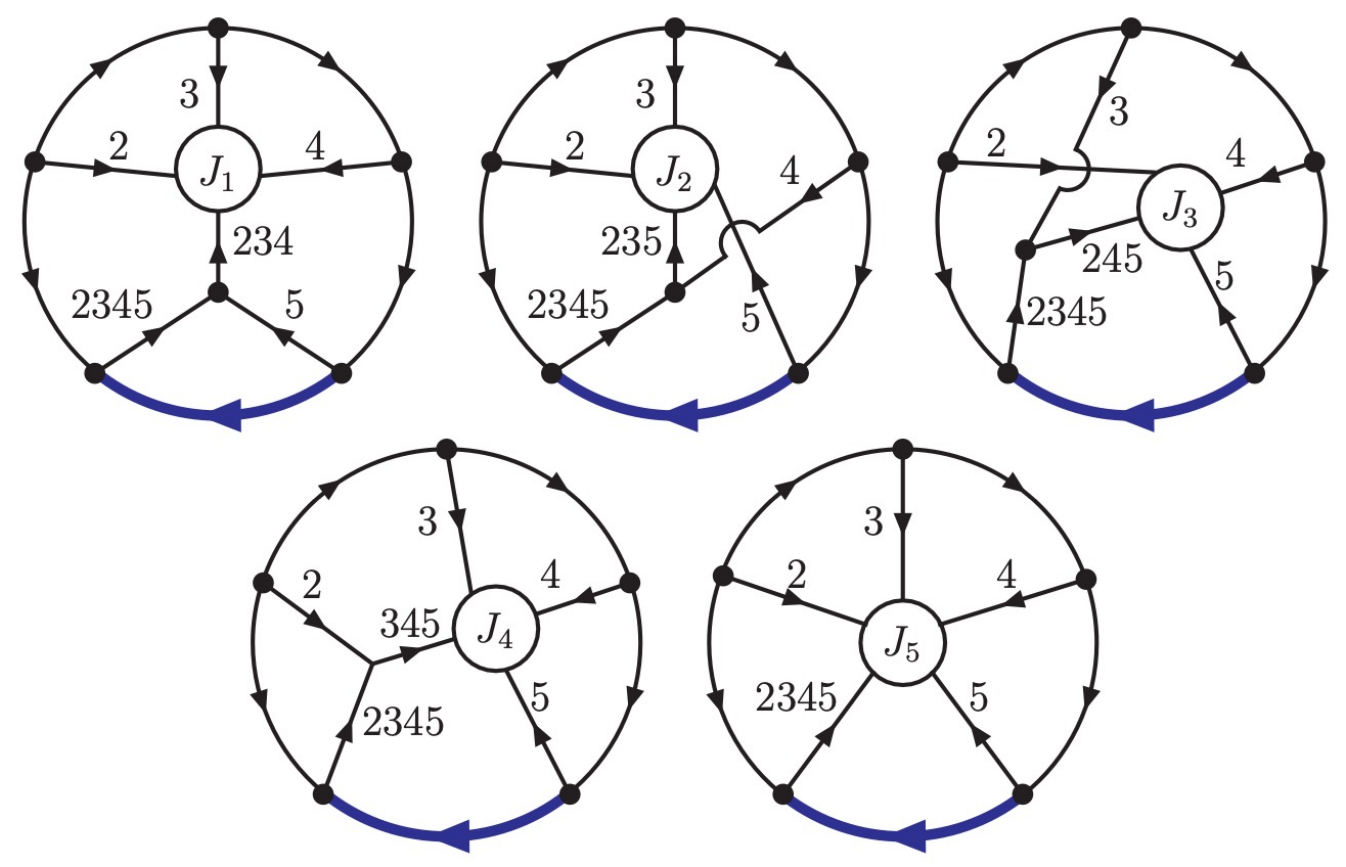}
\centering
\caption{Diagrammatic representation of the multiloop topologies associated to the sets defined in Tab.~\ref{tb:sets} where only the internal propagators are labeled. The blue bold line represents an arbitrary number of propagators. In the case of the channel $J_5$, the characterize sets are established by merging two of the common sets as depicted in Fig.\ref{fig:topo_J5}.
\label{fig:topologies5_Ji}}
\end{figure}

Inspired by the N$^4$MLT {\it universal topology}~\cite{Ramirez-Uribe:2020hes}, we recall the concept of a current to encode in a compact form the distinctive sets. This arrangement is diagrammatically represented in Fig.~\ref{fig:topologies5_J}.
Each specific topology is characterized by a pair of sets from the list $\{23,24,25,34,35,45,234,235,245,34\}$.
It is important to take into account that one of the internal sets, $2345$, is imposed by momentum conservation. The total number of distinctive pairs is fifteen, and they can be conveniently grouped in five channels as presented in Tab.~\ref{tb:sets} and displayed in Fig.~\ref{fig:topologies5_Ji}.
Many nonplanar topologies arise; in fact, they account for ten from the total of fifteen.

Each pair of sets from Tab.~\ref{tb:sets} is associated to a single Feynman diagram. The first $L$ common sets of propagators depend on one characteristic loop momentum $\ell_s$ and the momenta of their propagators have the form
\beq
q_{i_s}=\ell_s + k_{i_s} \qquad s\in \{1,\dots,L \}~.
\eeq
The remaining five common sets are denoted as linear combinations of all the loop momenta, explicitly:
\begin{align} \label{eq:momenta1}
q_{i_{(L+1)}}       & = - \sum_{s=1}^L \ell_s + k_{i_{(L+1)}} \, , \quad q_{i_{12}}    = -  \ell_1 - \ell_2 +k_{i_{12}} \, , \nn   \\
q_{i_{123}} & = - \sum_{s=1}^3 \ell_s +k_{i_{123}} \, , \quad 
q_{i_{1234}}    = - \sum_{s=1}^4 \ell_s   +k_{i_{1234}} \, , \nn \\
q_{i_{2345}} & = - \sum_{s=2}^5 \ell_s +k_{i_{2345}} \, ,
\end{align}
with $k_{i_s}, k_{i_{(L+1)}}, k_{i_{12}}, k_{i_{123}}, k_{i_{1234}}$ and $k_{i_{2345}}$ linear combinations of external momenta. 
\begin{table}[t]
\begin{center}
    \begin{tabular}{llll} \hline \hline 
$J_1$: \qquad & $\{ 234,23 \}$ \qquad\quad & $\{234,24\}~{\color{blue}\star}$  \qquad\quad & $\{ 234,34 \}$ \\ 
$J_2$: \qquad & $\{ 235,23 \}~{\color{blue}\star}$ \qquad\quad & $\{ 235,25 \}~{\color{blue}\star}$ \qquad\quad & $\{ 235,35 \}~{\color{blue}\star}$ \\
$J_3$: \qquad & $\{ 245,24 \}~{\color{blue}\star}$ \qquad\quad & $\{ 245,25 \}~{\color{blue}\star}$ \qquad\quad & $\{ 245,45 \}~{\color{blue}\star}$ \\ 
$J_4$: \qquad & $\{ 345,34 \}$ \qquad\quad & $\{ 345,35 \}~{\color{blue}\star}$ \qquad\quad & $\{ 345,45 \}$ \\ 
$J_5$: \qquad & $\{ 23,45 \}$ \qquad\quad & $\{ 24,35 \}~{\color{blue}\star}$ \qquad\quad & $\{ 25,34 \}~~{\color{blue}\star}$ \\
\hline \hline 
\end{tabular}
\end{center}
\caption{Pair of sets identifying the representative topologies of the N$^7$MLT family. The star indicates that the set is associated to a nonplanar diagram.} \label{tb:sets}
\end{table}
The extra sets are the distinctive key to generate each of the fifteen topologies. 
We identify the momenta of their propagators as different linear combinations of $\ell_2$, $\ell_3$, $\ell_4$ and $\ell_5$, writing them as
\begin{align}
q_{i_{rs}}  = -  \ell_r - \ell_s +k_{i_{rs}}~, \quad
q_{i_{rst}}  = -  \ell_r - \ell_s -  \ell_t +k_{i_{rst}} \, , \qquad r, s \in \{2,3,4,5\}\, .
\label{eq:momenta2} 
\end{align}
In order to achieve a single expression to represent the N$^7$MLT topologies, we merge the sets presented in Tab.~\ref{tb:sets} into a current labeled as $J$,
\beq \label{eq:current}
J=\cup_{i=1}^5 J_i~.
\eeq 
The setting of this scenario allows us to assemble the Feynman representation of the N$^7$MLT family as
\beq \begin{aligned} \label{eq:unified}
\mathcal{A}_{\rm N^7MLT}^{(L)} = \int_{\ell_1, \ldots, \ell_L } \mathcal{A}_F^{(L)}(1, \ldots, L+1, 12, 123, 1234, 2345, J)~,
\end{aligned} \eeq
diagrammatic represented in accordance with Fig.~\ref{fig:topologies5_J}.

The dual opening of Eq.~\eqref{eq:unified} is computed by the direct application of the nested residues through the LTD framework. 
With the purpose of obtaining a manageable dual expression, we also propose an ansatz based in a graphical interpretation of the opening.
The comparison between these two ingredients allow us to achieve a LTD representation exhibited in a factorized form in terms of simpler subtopologies,
\beq \begin{aligned} \label{eq:universal}
& \mathcal{A}_{\rm N^7MLT}^{(L)}(1, \ldots, L+1, 12, 123, 1234, 2345, J) \\
& = \mathcal{A}_{\rm N^6MLT}^{(5)}(1, \ldots, 5, 12, 123, 1234, 2345, J) \otimes \mathcal{A}_{\rm MLT}^{(L-5)}(6, \ldots, L+1)  \\
& + \mathcal{A}_{\rm N^5MLT}^{(4)}(1 \cup 2345, 2, 3, 4, 5 \cup 1234, 12, 123, J) \otimes \mathcal{A}_{\rm MLT}^{(L-4)}(\overline{6}, \ldots, \overline{L+1})~. 
\end{aligned} \eeq
The term ${\cal A}^{(L)}_{\rm N^{k-1}MLT}$ refers to the integrand of the corresponding topology in the LTD representation; integration over the $L$ loop momenta is assumed.
The convolution symbol indicates that each component is open independently, while the on-shell conditions of all elements act together over the remaining off-shell propagators. Regarding the selected on-shell propagators, they are restricted so as not to generate disjoint trees due to the dual opening.

\begin{figure}[t]
\includegraphics[scale=0.45]{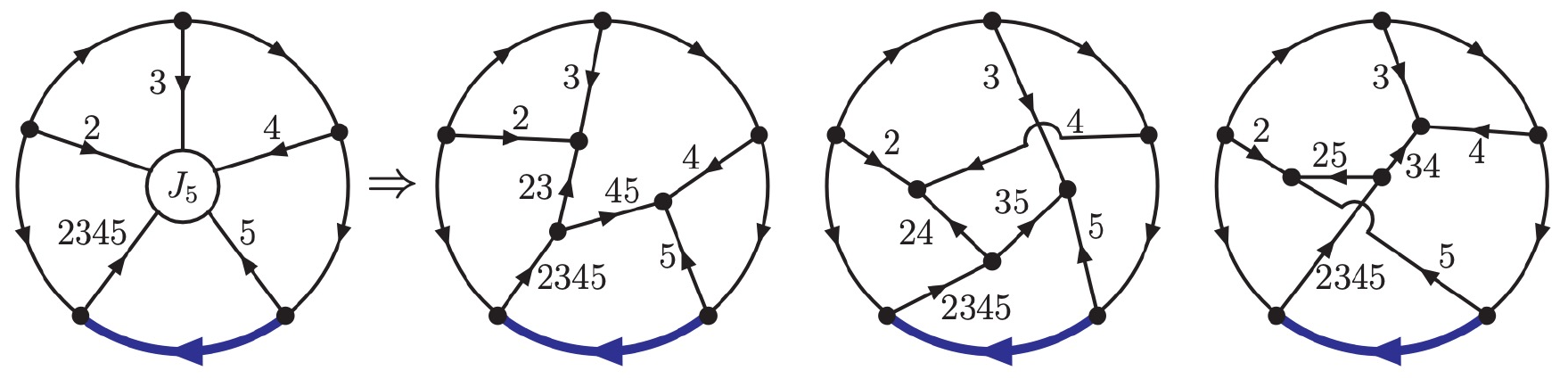}
\centering
\caption{Diagrammatic representation of the multiloop topologies conforming the channel $J_5$.
\label{fig:topo_J5}}
\end{figure}
\begin{figure}[t]
\includegraphics[scale=0.4]{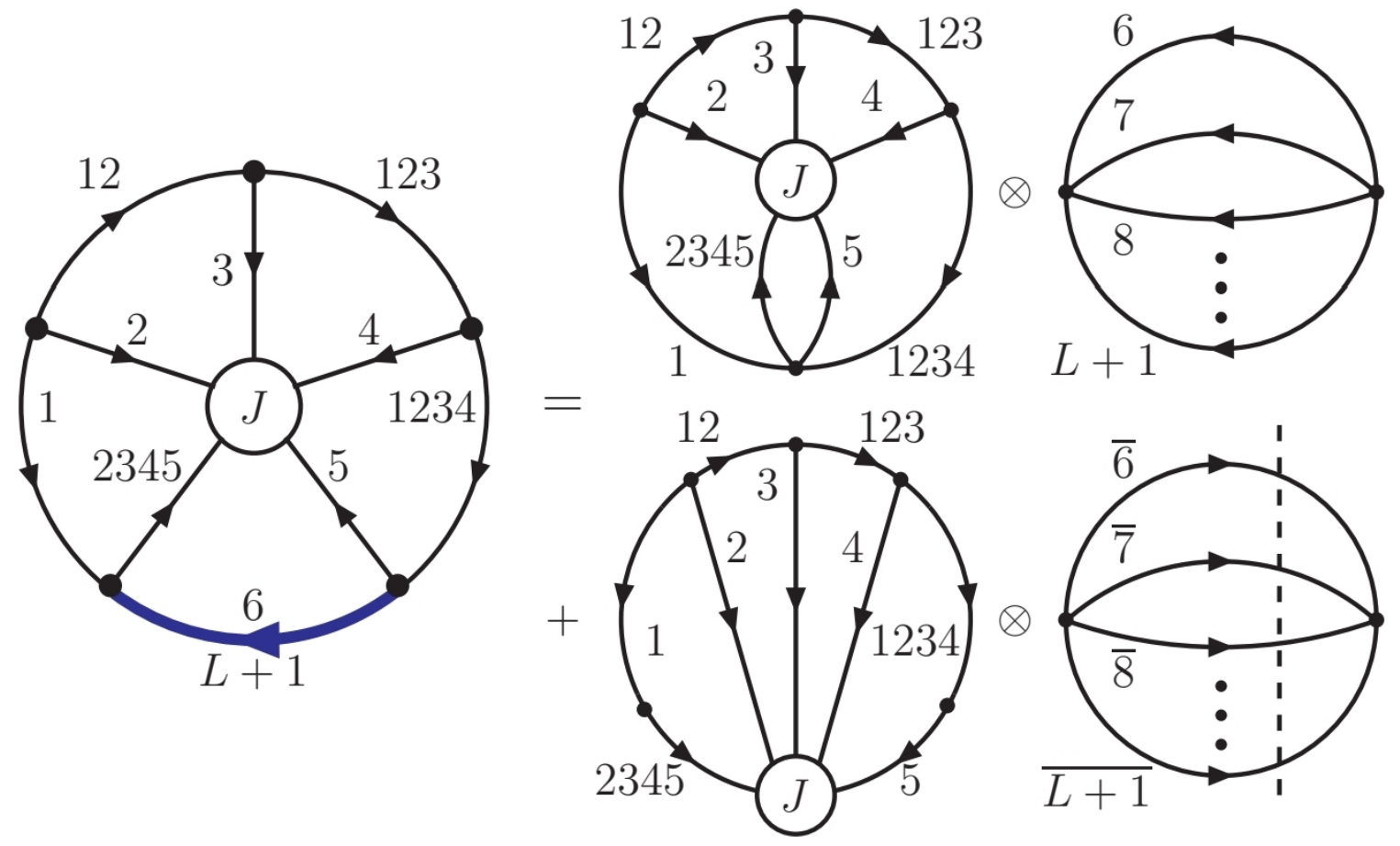}
\centering
\caption{Diagrammatic representation for the factorized opening of the multiloop N$^7$MLT general topology. Only the on-shell cut of the last MLT-like subtopology with reversed momentum flow is shown.
\label{fig:opening}}
\end{figure}

The LTD expression in Eq.~\eqref{eq:universal} opens any multiloop N$^7$MLT topology to connected trees, furthermore, it is valid for all N$^{k-1}$MLT configurations with $k\le 7$. 
The dual opening interpretation is depicted in Fig.~\ref{fig:opening}, where the diagram associated to the term $\mathcal{A}^{(5)}_{\rm N^6MLT}$ on the r.h.s. of Eq.~\eqref{eq:universal} contemplate all possible configurations with five on-shell propagators in the sets $\{1, 2, 3,4,12,123,1234,2345,J \}$, and the contribution of ${\cal A}^{(4)}_{\rm N^5MLT}$ in the second term assumes four on-shell conditions selected from $\{ 1\cup2345,2,3,4, 5\cup1234,12,123,J\}$. 
The term ${\cal A}_{\rm MLT}^{(L-5)} (6, \dots, L+1)$ is opened according to the MLT opening introduced in Ref.~\cite{Verdugo:2020kzh}; 
in ${\cal A}_{\rm MLT}^{(L-4)} (\overline 6, \dots, \overline {L+1})$ all the momentum flows are reversed and all the sets contain one on-shell propagator.  

The unfolding of Eq.~\eqref{eq:universal} is computed recursively through the subtopologies arising, for instance, the five-loop contribution opens as follows
\beq \begin{aligned} \label{eq:5Lsubtopology}
& \mathcal{A}_{\rm N^6MLT}^{(5)}(1, \ldots, 5, 12, 123, 1234, 2345, J) \\
& = \mathcal{A}_{\rm N^4MLT}^{(5)}(1, \ldots, 5, 12, 123, 1234, 2345)  \\
& + \sum_{s \in J} \mathcal{A}_{\rm N^4MLT}^{(4)}(1, \ldots, 5, 12, 123, 1234, 2345, {\bold s}) ~.
\end{aligned} \eeq
The first term in the r.h.s. gives a LTD contribution with all the propagators with momenta in $J$ remaining off shell.
The second term on the r.h.s. of Eq.~\eqref{eq:5Lsubtopology} considers all the contributions with a pair of propagators with momenta belonging to Tab.~\ref{tb:sets}, therefore, these dual trees correspond to a single representative channel. The bold symbol, ${\bold s}$, is used to identify the contributions with on-shell propagators belonging to $J$.

Concerning the four-loop subtopology in Eq.~\eqref{eq:universal}, the dual opening is given by
\beq \begin{aligned} \label{eq:4Lsubtopology}
& \mathcal{A}_{\rm N^5MLT}^{(4)}(1 \cup 2345, 2, 3, 4, 5 \cup 1234, 12, 123, J) \\
& = \mathcal{A}_{\rm N^3MLT}^{(4)}(1 \cup 2345, 2, 3, 4, 5 \cup 1234, 12, 123) \\
& + \sum_{s \in J} \mathcal{A}_{\rm N^3MLT}^{(3)}(1 \cup 2345, 2, 3, 4, 5 \cup 1234, 12, 123, {\bold s})~, \\
\end{aligned} \eeq
where similar to Eq.~\eqref{eq:5Lsubtopology}, the first term in the r.h.s. considers all the pair of propagators with momenta in $J$ off shell, whereas the second term is associated to a specific three-loop N$^3$MLT topology.  

Every representative topology in the N$^7$MLT family have common dual terms, those associated to the first term in the r.h.s. in Eqs.~\eqref{eq:5Lsubtopology} and \eqref{eq:4Lsubtopology} respectively,
\begin{align} \label{eq:common-terms}
\mathcal{A}_{\rm N^4MLT}^{(5)}(1, \ldots, 5, 12, 123, 1234, 2345) \quad \text{and} \quad \mathcal{A}_{\rm N^3MLT}^{(4)}(1 \cup 2345, 2, 3, 4, 5 \cup 1234, 12, 123)~. \end{align}
These contributions can be graphical represented with the first diagram in the r.h.s. of both sums depicted in Fig.~\ref{fig:opening} by replacing $J$ with a five point interaction among the internal sets, i.e., by taking $J$ as an empty set. 
To entirely unfold them and obtain the explicit dual terms, we continue to open the subtopologies recursively. Even if the ordering of opening any topology is arbitrary, following the guide of a diagrammatic interpretation allow us to work in a more manageable form. 

Concerning the common dual terms in the N$^7$MLT topologies, we have that
\beq \label{eq:common4L}
\mathcal{A}_{\rm N^3MLT}^{(4)}(1 \cup 2345, 2, 3, 4, 5 \cup 1234, 12, 123)
\eeq
opens according Ref.~\cite{Ramirez-Uribe:2020hes} by considering the set $234$ as empty and, in the case of the five-loop contribution, the explicit dual opening is given by
\beq \begin{aligned} \label{eq:common5L}
& \mathcal{A}_{\rm N^4MLT}^{(5)}(1,\ldots,5,12,123,1234,2345)  \\
=& \mathcal{A}_{\rm N^4MLT}^{(4)}(1,2,3,4,12,123,1234)\otimes\mathcal{A}_{\rm MLT}^{(1)} (5, 2345) \\
&+ \left[\mathcal{A}_{\rm NMLT}^{(3)}(1,2,3,12,123) + \left(\mathcal{A}_{\rm MLT}^{(2)}(1,2,12) + \mathcal{A}_{\rm MLT}^{(2)}(1,\overline{3})\right)\right.\\
&\otimes \left.\mathcal{A}_{\rm MLT}^{(1)}(\overline{4})\right]\otimes \mathcal{A}_{\rm MLT}^{(2)}(\overline{5},\overline{2345}) + \left[ \mathcal{A}_{\rm MLT}^{(2)}(\overline{1234},2\cup\overline{12})\otimes \mathcal{A}_{\rm MLT}^{(1)}(3,123) \right.\\
&+ \mathcal{A}_{\rm MLT}^{(3)}(\overline{1234},\overline{3},\overline{123}) + \mathcal{A}_{\rm MLT}^{(2)}(\overline{1234},2\cup\overline{3} \overline{12}) \otimes \mathcal{A}_{\rm MLT}^{(1)}(\overline{4}) \\
&+ \left.\left(\mathcal{A}_{\rm MLT}^{(2)}(3,\overline{12})+\mathcal{A}_{\rm MLT}^{(2)}(\overline{1234},2\cup\overline{3}\cup\overline{12}) \right)\otimes\mathcal{A}_{\rm MLT}^{(1)}(4) \right] \otimes \mathcal{A}_{\rm MLT}^{(2)}(5, 2345)~.
\end{aligned} \eeq
The number of dual terms arising from the contributions in Eqs.~\eqref{eq:common4L} and \eqref{eq:common5L} for an arbitrary number of loops is given by $11(6L-19)$.

The contributions corresponding to the second term in the r.h.s. in Eqs.~\eqref{eq:5Lsubtopology} and \eqref{eq:4Lsubtopology} depend on the particular topology to be opened. In the following section we take a specific example and present the explicit contributions characterising the topology.

\section{Specific channels}
\label{sec:Examples}
\begin{figure}[t]
\includegraphics[scale=0.4]{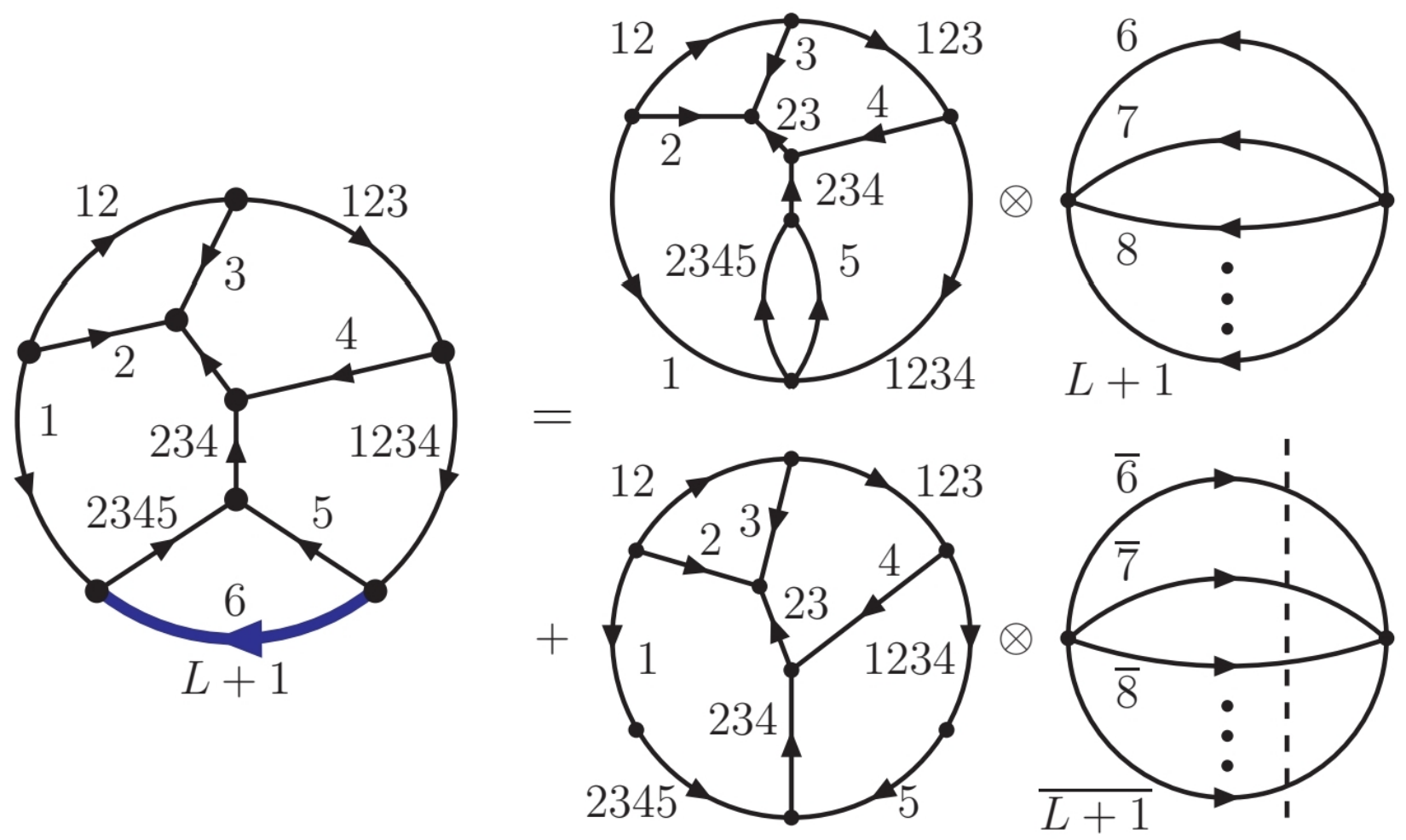}
\centering
\caption{Diagrammatic representation for the topology having as characterized set $\{234,23 \}$ and its factorized dual opening.
\label{fig:topo_234}}
\end{figure}
In order to deepen in the dual opening methodology, we present the development of the topology associated to the pair of sets $\{234,23\}$ belonging to the current $J_1$. The factorized dual opening is obtained through the application of Eq.~\eqref{eq:universal} replacing $J$ by $\{234,23\}$. The diagrammatic representation of the associated topology and its factorized dual opening is depicted in Fig.~\ref{fig:topo_234}.

\begin{figure}[t]
\includegraphics[scale=0.40]{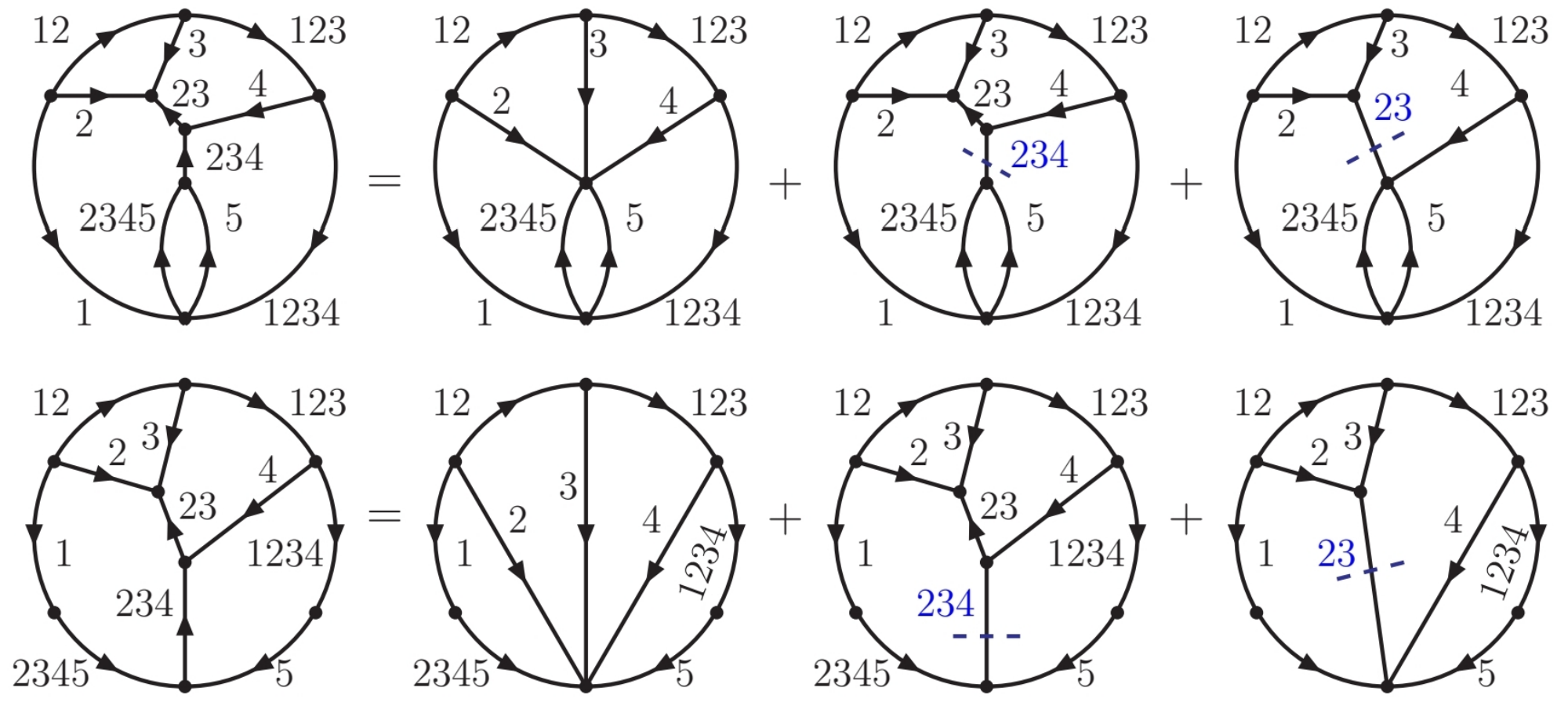}
\centering
\caption{Diagrammatic representation of the sum of contributions associated to the five-loop N$^6$MLT (top) and the four-loop N$^4$MLT (bottom) dual opening.
\label{fig:topo45L}}
\end{figure}

The common terms in the fifteen topologies are the ones associated to Eqs.\eqref{eq:common4L} and \eqref{eq:common5L} which are illustrated with the first diagram in the r.h.s. of each sum in Fig.~\ref{fig:topo45L}.
From the previous section we know that the contributions for this specific topology arise from the second term in the r.h.s. of Eqs.~\eqref{eq:5Lsubtopology} and \eqref{eq:4Lsubtopology} which are diagrammatically represented by the sum of the second and third diagrams in the r.h.s. of each sum in Fig.~\ref{fig:topo45L}.

The unfolding of the second term in the r.h.s. of Eq.~\eqref{eq:5Lsubtopology} with $s\in\{234,23\}$ is given by
\beq\begin{aligned} \label{eq:234-23-5L}
& \mathcal{A}_{\rm N^4MLT}^{(4)}(1, \ldots, 5, 12, 123, 1234, 2345,23, {\color{blue}234})  \\
& + \mathcal{A}_{\rm N^4MLT}^{(4)}(1, \ldots, 5, 12, 123, 1234, 2345, {\color{blue}23})  \\
=& \left[\left(\mathcal{A}_{\rm N^2MLT}^{(3)}(1,2,3,12,123,23) + \mathcal{A}_{\rm {MLT}}^{(2)} (1,2,12)\otimes\mathcal{A}_{\rm MLT}^{(1)}(\overline{4}) + \mathcal{A}_{\rm MLT}^{(3)}(1,\overline{3},\overline{4})\right) \right.\\
&\otimes\mathcal{A}_D^{(1)}(\overline{234}) + \left(\mathcal{A}_{\rm MLT}^{(2)}(\overline{1234},2\cup\overline{12})\otimes\mathcal{A}_{\rm MLT}^{(1)}(3,4\cup{23}) + \mathcal{A}_{\rm MLT}^{(3)} (\overline{1234},\overline{3},\overline{4}\cup\overline{23}) \right. \\
&+ \left.\left.\mathcal{A}_{\rm MLT}^{(3)}(12,3,4) + \mathcal{A}_{\rm MLT}^{(3)} (4\cup\overline{1234},2\cup\overline{3}\cup\overline{12},\overline{123})\right) \otimes\mathcal{A}_D^{(1)}(234) \right]\otimes\mathcal{A}_{\rm MLT}^{(1)}(5,2345) \\
+& \left[ \left(\mathcal{A}_{\rm MLT}^{(2)}(1,2,12) + \mathcal{A}_{\rm MLT}^{(2)}(1,\overline{3})\right)\otimes\left(\mathcal{A}_{\rm MLT}^{(2)}(4,5,2345)+\mathcal{A}_{\rm MLT}^{(1)}(5,2345)\otimes\mathcal{A}_{\rm MLT}^{(1)}(\overline{1234})\right) \right.\\
& + \left. \mathcal{A}_{\rm MLT}^{(1)}(2\cup \overline{3}, 12)\otimes \mathcal{A}_{\rm MLT}^{(2)}(5, 2345)\otimes \mathcal{A}_{\rm MLT}^{(1)}(\overline{1234}) \right] \otimes\mathcal{A}_D^{(1)}(\overline{23}) \\
+&\left[ \left(\mathcal{A}_{\rm MLT}^{(2)}(\overline{123},\overline{3},12) + \mathcal{A}_{\rm MLT}^{(2)}(\overline{123},2)\right)\otimes\mathcal{A}_{\rm MLT}^{(2)}(4,5,2345) + \left( \mathcal{A}_{\rm MLT}^{(2)}(\overline{4}\cup \overline{123},\overline{3},12) \right.\right.\\
&+ \left.\left.\mathcal{A}_{\rm MLT}^{(2)}(\overline{4} \cup\overline{123},2)\right) \otimes \mathcal{A}_{\rm MLT}^{(1)}(5,2345)\otimes\mathcal{A}_{\rm MLT}^{(1)}(\overline{1234}) \right] \otimes\mathcal{A}_D^{(1)}(23)~,
\end{aligned}\eeq
and the one corresponding to Eq.~\eqref{eq:4Lsubtopology},
\beq \begin{aligned} \label{eq:234-23-4L}
&\mathcal{A}_{\rm N^3MLT}^{(3)}(1 \cup 2345, 2, 3, 4, 5 \cup 1234, 12, 123,23, {\color{blue}234}) \\
& + \mathcal{A}_{\rm N^3MLT}^{(3)}(1 \cup 2345, 2, 3, 4, 5 \cup 1234, 12, 123, {\color{blue}23}) \\
& =\left[ \mathcal{A}_{\rm N^2MLT}^{(3)}(1 \cup 2345, 2, 3, 12, 123, 23) + \mathcal{A}_{\rm MLT}^{(2)}(1 \cup 2345, 2, 12) \otimes \mathcal{A}_{\rm MLT}^{(1)} (\overline{4}) \right. \\
& +\left. \mathcal{A}_{\rm MLT}^{(3)}(1 \cup 2345, \overline{3}, \overline{4}) \right] \otimes  \mathcal{A}_D^{(1)}(\overline{234}) \\
&+\left[ \mathcal{A}_{\rm MLT}^{(2)}(\overline{5} \cup \overline{1234}, 2 \cup \overline{12}) \otimes \mathcal{A}_{\rm MLT}^{(1)}(3, 4 \cup 23) + \mathcal{A}_{\rm MLT}^{(3)}(\overline{5}\cup \overline{1234}, \overline{3}, \overline{4}\cup \overline{23})\right. \\
& +\left. \mathcal{A}_{\rm MLT}^{(3)}(\overline{12}, 3, 4) + \mathcal{A}_{\rm MLT}^{(3)}(\overline{5}\cup \overline{1234}\cup \overline{4}, 2 \cup \overline{3}\cup \overline{12}, \overline{123}) \right]  \otimes  \mathcal{A}_D^{(1)}(234) \\
&+ \left(\mathcal{A}_{\rm MLT}^{(2)}(1\cup 2345, 2, 12) + \mathcal{A}_{\rm MLT}^{(2)}(1\cup 2345, \overline{3}) \right) \otimes \mathcal{A}_{\rm MLT}^{(1)}(4, 5\cup 1234) \otimes  \mathcal{A}_D^{(1)}(\overline{23}) \\
&+ \left[ \left( \mathcal{A}_{\rm MLT}^{(2)}(\overline{123}, \overline{3}, 12)+\mathcal{A}_{\rm MLT}^{(2)}(\overline{123},2) \right) \otimes \mathcal{A}_{\rm MLT}^{(1)}(4, 5\cup 1234) \right. \\
&+ \left. \mathcal{A}_{\rm MLT}^{(1)}(2\cup \overline{3}, 12) \otimes \mathcal{A}_{\rm MLT}^{(2)}(\overline{4}, \overline{5}\cup \overline{1234}) \right] \otimes  \mathcal{A}_D^{(1)}(23)~.
\end{aligned} \eeq
The sets in blue in the arguments of the terms in the l.h.s. of Eqs.~\eqref{eq:234-23-5L} and \eqref{eq:234-23-4L} indicate the set that we use as a starting cut in the second and third diagram of each sum in Fig.~\ref{fig:topo45L}. After this action we end up with known subtopologies that we recursively compute by applying the results from Refs.~\cite{Verdugo:2020kzh,Ramirez-Uribe:2020hes}.
The total number of dual terms emerging from Eqs.~\eqref{eq:234-23-5L} and \eqref{eq:234-23-4L} for an arbitrary number of loops is given by $119L-380$. The number of dual terms arising from every specific channel for an arbitrary number of loops is summarized in Tab.~\ref{tb:counting}.
\begin{table}[ht]
\begin{center}
    \begin{tabular}{llll} \hline \hline 
$J_1$: \qquad & $~~~\{ 234,23 \}~~119~L-380$ \qquad & ${\color{blue}\star}~\{234,24\}~~132~L-418$  \qquad & $~~~\{ 234,34 \}~~119~L-380$ \\ 
$J_2$: \qquad & ${\color{blue}\star}~\{ 235,23 \}~~143~L-473$ \qquad & ${\color{blue}\star}~\{ 235,25 \}~~158~L-519$ \qquad & ${\color{blue}\star}~\{ 235,35 \}~~165~L-554$ \\
$J_3$: \qquad & ${\color{blue}\star}~\{ 245,24 \}~~165~L-554$ \qquad & ${\color{blue}\star}~\{ 245,25 \}~~165~L-554$ \qquad & ${\color{blue}\star}~\{ 245,45 \}~~165~L-583$ \\ 
$J_4$: \qquad & $~~~\{ 345,34 \}~~143~L-500$ \qquad & ${\color{blue}\star}~\{ 345,35 \}~~165~L-583$ \qquad & $~~~\{ 345,45 \}~~158~L-575$ \\ 
$J_5$: \qquad & $~~~\{ 23,45 \}~~~~143~L-500$ \qquad & ${\color{blue}\star}~\{ 24,35 \}~~~~165~L-554$ \qquad & ${\color{blue}\star}~\{ 25,34 \}~~~~143~L-473$ \\
\hline \hline 
\end{tabular}
\end{center}
\caption{Numbers of dual terms for the individual channels.} \label{tb:counting}
\end{table}

It is worth mentioning that among the fifteen configurations some of them are topologically equivalent. For example, the dual representation of the diagram characterized by $\{234,34\}$ can be obtained from the dual representation computed to the topology characterized with $\{234,23\}$, by the substitution $23\rightarrow 34$ and the exchange $2\leftrightarrow 4$, 

\beq
\mathcal{A}_{\rm N^7MLT}^{(L)}(1, \ldots, 234,23) \quad \xrightarrow[2\leftrightarrow 4]{23\rightarrow 34} \quad
\mathcal{A}_{\rm N^7MLT}^{(L)}(1, \ldots, 234,34)~.  \nn
\eeq
For the remaining equivalent topologies, we require to exchange $2\leftrightarrow 4$ and $5\leftrightarrow 2345$, as well as applying the proper substitution between the distinctive sets. Explicitly, we have that
\begin{align} 
&\mathcal{A}_{\rm N^7MLT}^{(L)}(1, \ldots, 235,23) \quad \xrightarrow[2\leftrightarrow 4,\; 5\leftrightarrow 2345]{235\rightarrow 25,\;23\rightarrow 34} \quad
\mathcal{A}_{\rm N^7MLT}^{(L)}(1, \ldots, 25,34)~,  \nn \\
&\mathcal{A}_{\rm N^7MLT}^{(L)}(1, \ldots, 235,35) \quad \xrightarrow[2\leftrightarrow 4,\; 5\leftrightarrow 2345]{235\rightarrow 25,\;35\rightarrow 245} \quad
\mathcal{A}_{\rm N^7MLT}^{(L)}(1, \ldots, 245,25)~,  \nn \\ 
&\mathcal{A}_{\rm N^7MLT}^{(L)}(1, \ldots, 245,24) \quad \xrightarrow[2\leftrightarrow 4,\; 5\leftrightarrow 2345]{245\rightarrow 35} \quad
\mathcal{A}_{\rm N^7MLT}^{(L)}(1, \ldots, 24,35)~,  \nn \\ 
&\mathcal{A}_{\rm N^7MLT}^{(L)}(1, \ldots, 245,45) \quad \xrightarrow[2\leftrightarrow 4,\; 5\leftrightarrow 2345]{245\rightarrow 35,\;45\rightarrow 345} \quad
\mathcal{A}_{\rm N^7MLT}^{(L)}(1, \ldots, 345,35)~,  \nn \\ 
&\mathcal{A}_{\rm N^7MLT}^{(L)}(1, \ldots, 345,34) \quad \xrightarrow[2\leftrightarrow 4,\; 5\leftrightarrow 2345]{345\rightarrow 45,\;34\rightarrow 23} \quad
\mathcal{A}_{\rm N^7MLT}^{(L)}(1, \ldots, 23,45)~. \nn 
\end{align}

Certainly, there are topologies that cannot be obtained from a direct replacement of indices, mainly those related with non planar topologies. Nevertheless, the direct application of the LTD shows that there are no disjoint trees in the forest of the N$^7$MLT topology and these feature is important to compute physical observables at high precision. Therefore, we let the application of these expressions to further analysis.

\section{Conclusions and outlook}
\label{sec:conclusions}
In this work, we applied the Loop-Tree Duality (LTD) and the nested residue representation to characterize all the possible five-loop Feynman diagrams. Following the approach presented in Ref. \cite{Ramirez-Uribe:2020hes}, we identified a \emph{universal topology}, i.e. the N$^7$MLT diagram, and then related the possible subtopologies through factorization formulae. As already known for simpler cases (such as N$^4$MLT, N$^3$MLT and N$^2$MLT, studied in Refs. \cite{Verdugo:2020kzh,Aguilera-Verdugo:2020kzc,Ramirez-Uribe:2020hes}), these factorization formulae lead to a recursive representation of complex topologies in terms of simpler ones. In this way, it is possible to understand the singular properties of all the five-loop amplitudes in terms of objects with a lower complexity.

Beyond the important recursive relations found among multiloop amplitudes, the developments shown in this paper could be further explored to simplify the calculation of important quantities at five-loop accuracy. In this way, the LTD representation and its factorization properties could allow to break the bottleneck of multiloop multileg calculations in QFT.

\section*{Acknowledgements}
SRU acknowledges support from MCIN/AEI/10.13039/501100011033, Grant No. PID2020-114473GB-I00, Consejo Nacional de Ciencia y Tecnología and Universidad Aut\'onoma de Sinaloa. The work of R.~J.~Hern\'andez-Pinto is supported by CONACyT (Mexico) Project No. 320856 (Paradigmas y Controversias de la Ciencia 2022), Ciencia de Frontera 2021-2042 and Sistema Nacional de Investigadores as well as by PROFAPI 2022 Grant No. PRO\_A1\_024 (Universidad Aut\'onoma de Sinaloa). GS is partially supported by Programas Propios II (Universidad de Salamanca), EU Horizon 2020 research and innovation program STRONG-2020 project under grant agreement No. 824093 and MSCA H2020 USAL4EXCELLENCE-PROOPI-391 cofund project under grant agreement No 101034371.


\bibliographystyle{JHEP}
\bibliography{refs}

\end{document}